\documentclass[11pt]{article}
\usepackage{graphicx}
\usepackage{grffile}
\usepackage{epstopdf}
\usepackage[titletoc]{appendix}
\usepackage[font=small, labelfont={small,bf}, margin=0.5cm]{caption}

\title{Constraints on Chronologies}

\author{Alfred Shapere$^1$ and Frank Wilczek$^2$\\\\
  \small\it $^1$Department of Physics and Astronomy, University of Kentucky,\\  \small\it Lexington KY 40506 USA\\
  \small\it $^2$ Center for Theoretical Physics, Department of Physics,\\   \small\it MIT, Cambridge MA 02139 USA}

\begin{document}

\maketitle

\begin{abstract}
The time ordering of two spacelike separated events is arbitrary, when all inertial frames are taken into account, but for three or more events it is not generally so.   We determine the structure of possible time orderings, or chronologies, for multiple events in any number of dimensions, analytically and exhaustively for three events in four space-time dimensions, algorithmically in other cases.  We also formulate an alternative criterion, based on convexity, for determining the allowed chronologies of a set of events. We show how the metric of a Lorentz invariant spacetime can be partially reconstructed from a knowledge of the chronologies it supports. Finally, we propose a different but related criterion for allowed chronologies in curved spacetimes.
\end{abstract}

\medskip
Special relativity introduces the idea that many alternative definitions of time, appropriate to different inertial frames, have equal standing.  Each frame introduces a chronology of spacetime events; i.e., two given events $A, B$ are assigned times $t_A, t_B$. 
Famously: If $A$ and $B$ are timelike or lightlike separated, {\it i.e.}, $(A-B)^2 \geq 0$ in signature $({+}\,{-}\,{-}\,{-})$, then the same time ordering, say $t_A < t_B$, obtains in every inertial frame; but if $A$ and $B$ are spacelike separated, $(A-B)^2 < 0$, one finds $t_A < t_B$ in some frames while $t_A > t_B$ in others.  It is natural to ask what happens when one considers more than two spacelike separated events.  Do all possible time orderings occur, in some frame or other?  If not, what are the restrictions?   Here we investigate this question from several perspectives, and find that the answer is far from trivial.  

We begin with a detailed investigation of the case of three spacelike separated events $A$, $B$, and $C$ in four dimensions, for which we will find a simple criterion for whether or not the events can be placed in arbitrary time order, by appropriate choices of reference frames.   The criterion can be stated in several ways, and boils down to the statement that all six possible orderings can be achieved if and only if the two-dimensional plane passing through the three events contains no timelike vectors.  If this plane does contain a timelike vector, then typically at least two orderings will be excluded, implying, for example, that $A$ may not precede both $B$ and $C$.  On the other hand, if no such timelike vector exists, then a timelike vector normal to the plane passing through $A$, $B$, and $C$ necessarily exists.  In a reference frame for which this normal vector is parallel to the time direction, 
$A$, $B$, and $C$ are simultaneous, and by making small Lorentz transformations all six orderings can be realized.

Such considerations are relevant in the conventional formulation of quantum mechanics, in which   
surfaces of simultaneity play a special role, since the state of the world is specified on such surfaces.   If $A, B, C$ are mutually spacelike separated events, then different observers may use wave functions where $A$ precedes $B$, so that measurement at $A$ collapses the wave function accessible to $B$; or alternatively where $B$ precedes $A$, so that measurement at $B$ collapses the wave function accessible to $A$, and similarly for the other pairs.   With the chronology constraint that $A$ can never precede both $B$ and $C$, for example, we have the peculiar situation that measurement at $A$ can cause collapse at either $B$ or $C$, in different frames, but never both.   That peculiarity does not present any causal paradoxes, but it does emphasize the conventional, and perhaps artificial, nature of wave functions and, especially, collapse.   (We plan to elaborate further on the implications of chronology conditions for the logic of quantum theory, which initially motivated the present investigation, in future work.) 

Next we generalize this result to arbitrarily many events (not necessarily spacelike separated) in arbitrarily many dimensions, following an approach due to  Stanley \cite{Stanley} and Heiligman \cite{Heiligman}.  Their approach is based on the observation that in the space of inertial frames parameterized by boost velocities, any two events define a hyperplane of simultaneity, a linear space of codimension one along which the two events are simultaneous.  A set of $k$ events thus determines $\frac{1}{2}k(k-1)$ hyperplanes of simultaneity, which can be analyzed using methods of the theory of hyperplane arrangements \cite{hyperplanes}.  We present a general algorithm for determining all possible time-orderings of $k$ events in $n+1$ dimensions.

In Section 3 we take a different, complementary approach and formulate a purely geometric criterion, based on convexity, that answers the question of when a time-ordering exists in which a given collection of events precedes another. 

In the following section we pose the question: Given a knowledge of all allowed chronologies, to what extent can we reconstruct the metric?  For flat spacetimes, we argue that the metric can be reconstructed up to a constant scale factor, and for curved spacetimes, up to a conformal factor. 

Finally we study time-orderings in curved space, where a different approach is required.   We formulate a general criterion, applicable to strongly causal spacetimes, for determining whether a time-slicing exists that realizes a given chronological relationship between two sets of events.

\bigskip

\section{Three Events}

Consider three mutually spacelike separated events $A$, $B$, $C$ in 3+1 dimensions.  We will derive a quantitative criterion that will turn out to distinguish when these events can be placed in an arbitrary time order and when they cannot. 

Define the spacelike 4-vectors $s_1 = A-C$, $s_2 = B-C$, and consider the plane consisting of all real linear combinations $\lambda_1 s_1 + \lambda_2 s_2$.  Although this plane is spanned by two spacelike vectors, it may contain vectors that are not spacelike \cite{Robb}. To develop a criterion for when this situation occurs, it is useful to consider the quantity $q = (\lambda_1 s_1 + \lambda_2 s_2)^2$, which may be viewed as a quadratic form in two real variables $\lambda_1, \lambda_2$.  $q(\lambda_1,\lambda_2)$, which is negative when either $\lambda_1$ or $\lambda_2$ vanishes, will fail to be  everywhere negative if and only if the equation $q =0$ has a solution.  This gives, in effect, a quadratic equation in $\lambda_1/\lambda_2$.    That equation has real roots if and only if 
\begin{equation}\label{criterionOne}
(s_1 \cdot s_2 )^2 \, \geq  \, s_1^2 s_2^2.
\end{equation}
We shall call this the chronology criterion.  From the derivation, equality holds in Eqn.~(\ref{criterionOne}) if and only if the plane spanned by $s_1, s_2$ contains lightlike vectors, but not timelike ones.  
We will show that when (and only when) this criterion is met, there are non-trivial constraints on the possible chronologies of $A$, $B$, and $C$.  

The chronology criterion can be expressed in several forms.   If we put $s_1 \equiv (t_1, \vec x_1), s_2 \equiv (t_2, \vec x_2)$, some simple algebra yields the explicit, elementary form:
\begin{equation}\label{criterionTwo}
(t_1 \vec x_2 - t_2 \vec x_1)^2 \, \geq \, \vec x_1^2 \vec x_2^2 - (\vec x_1 \cdot \vec x_2)^2.
\end{equation}
Alternatively, if we consider a frame especially adapted to $v_1, v_2$, in which
\begin{eqnarray}
v_1  &=&  |v_1| \cdot (0, 1, 0, 0) \nonumber \\
v_2  &=&  |v_2| \cdot (\sinh \eta, \cosh \eta \cos \theta, \cosh \eta \sin \theta, 0) ,
\end{eqnarray}
then the chronology criterion becomes
\begin{equation}
| \cosh \eta \,\cos \theta \, | \geq 1 .
\end{equation}
Thus the chronology criterion is always satisfied when $\theta=0$, i.e. the three points are spatially collinear, and never satisfied when $\theta=\pi/2$.\footnote{Note that in 1+1 dimensions the right-hand side of Eqn.~(\ref{criterionTwo}) vanishes, and the criterion is always met.}  

Eqn.~(\ref{criterionOne}) in any of these forms hides its symmetry among $A$, $B$, $C$.   By expanding it out and expressing everything in terms of standard invariants, we get a manifestly symmetric expression:
\begin{eqnarray}\label{criterionThree}
&(AB)^2 - A^2B^2 + (BC)^2 - B^2 C^2 + (AC)^2 - A^2C^2 \nonumber \\
&+ 2 (AB) C^2 + 2 (BC) A^2 + 2 (AC) B^2 \\
&- 2 (AB)(BC) - 2 (BC)(AC) - 2 (CA)(AB) \, \geq \, 0 . \nonumber
\end{eqnarray}
We will give another, more geometric derivation of this version of the criterion at the end of this section.

\bigskip

\noindent{\it Case One: Criterion Met}

\bigskip

If the chronology criterion is met,  then we can find $\lambda_1$, $\lambda_2$ such that 
\begin{equation}\label{interpretationOne}
\lambda_1 v_1 + \lambda_2 v_2 \, = \, {\rm lightlike ~ or~ timelike}
\end{equation}
In terms of $A, B, C$, we can express this condition in the manifestly symmetric form
\begin{eqnarray}\label{interpretationTwo}
a A + b B + c C \, &=& \, {\rm lightlike~ or ~  timelike} \nonumber \\
a + b + c \, &=& \, 0 .
\end{eqnarray}

In Eqn.~(\ref{interpretationTwo}), without significant loss of generality we may assume $a, b > 0$, $c < 0$.  Since the 4-vector $aA+bB+cC=a(A-C) + b (B-C)$ is timelike or lightlike, the sign of its time coordinate is the same in all frames.  Let us say it is positive.   Then it can never be the case that $C$ comes after {\it both\/} $A$ and $B$; for if $(A-C)^t < 0$ {\it and\/} $(B-C)^t < 0$ we would have $a(A-C)^t + b(B-C)^t <0$, contrary to hypothesis. 
$C$ can still occur either after $A$ or after $B$ (though not both), so there must exist frames in which the chronology is 
$A^t < C^t  < B^t $ and others in which it is $B^t < C^t < A^t$.   Because the space of boosts is connected, and the mapping from it onto time orderings of $A, B, C$ is continuous, we must be able to interpolate between possible orderings through swaps of adjoining letters.   Thus the intermediate orderings $C^t < B^t < A^t$ and $C^t < A^t < B^t$ will also occur.   (Note that all three events cannot be simultaneous in any frame.)  

In the degenerate case when $A$, $B$, $C$ lie on a line,  one of them, say $B$, lies between the others.    In this case only the time orderings $A^t < B^t < C^t$ or $C^t < B^t < A^t$, or complete equality, are possible.   

Summing up: If the criterion Eqn.~(\ref{criterionThree}) is met, then the possible chronologies among the events $A, B, C$ are restricted in an interesting way.   Generically, either there is one event that cannot occur before both the others, or there is one event that cannot occur after both the others; within either alternative, all chronologies consistent with the stated constraint occur in some frame.

\bigskip

\noindent{\it Case Two: Criterion Unmet}

\bigskip

If the criterion (\ref{criterionOne}) is not met, then the plane spanned by $A-C$, $B-C$ consists entirely of spacelike vectors.   We can choose a frame such that a timelike vector normal to this plane defines the direction of time.  In that frame, $A$, $B$, $C$ are simultaneous.   By making small hyperbolic rotations about the axis through $A$ and $B$ (i.e. boosts in directions perpendicular to $A-B$) we can make $C$ occur slightly earlier or later, without changing the time coordinates of $A$ and $B$.   Combining operations of this kind, clearly we can find frames in which $A$, $B$, $C$ occur in any chronological order.    

\bigskip

\noindent{\it Geometric Formulation}

\bigskip

As we proceed, a more geometric formulation of the chronology criterion will prove useful.  Satisfaction of the criterion (\ref{criterionOne}) implies that the subspace spanned by $s_1$ and $s_2$ contains timelike or lightlike vectors. Because any vector orthogonal to a timelike vector must be spacelike, it follows that the subspace orthogonal to $s_1$ and $s_2$ must consist entirely of spacelike vectors.  (We leave the lightlike case to the reader.) Let $s_3$ and $s_4$ be two mutually orthogonal 4-vectors spanning this subspace; then 
\begin{equation}\label{wedges}
s_3 \wedge s_4 \ \propto\  *(s_1 \wedge s_2)
\end{equation} 
Squaring both sides gives
\begin{equation}\label{newcriterion}
s_3^2 s_4^2 \ \propto\  (s_1\cdot s_2)^2 - s_1^2 s_2^2
\end{equation} 
since we have assumed that $s_3\cdot s_4 = 0$.  The proportionality constant after squaring is positive, so the expressions on both sides of the last equation have the same sign.  It follows that if $s_3$ and $s_4$ are both spacelike, then the expression on the right side is strictly positive -- this is precisely the chronology criterion.  Conversely if the chronology criterion is satisfied then $s_3$ and $s_4$ must both be spacelike, since, being orthogonal, they cannot both be timelike. 

These considerations suggest a way of rewriting the symmetric version of the chronology criterion, Eqn.~(\ref{criterionThree}), 
in a more compact form. First express 
the right side of Eqn.~(\ref{wedges}) in terms of $A$, $B$, and $C$: 
\begin{equation}
*(s_1 \wedge s_2) = *[(A-C)\wedge(B-C)]=*(A\wedge B + B\wedge C + C\wedge A),
\end{equation} 
then square to obtain the chronology criterion in the form
\begin{equation}\label{criterionFour}
(A\wedge B + B\wedge C + C\wedge A)^2 \, \le \, 0 
\end{equation} 
This expression has the nice feature that although it involves the events themselves, as opposed to their differences, it is manifestly invariant under a common translation $A, B, C \, \rightarrow \, A + d, B +d , C +d$.
Expanded out, Eqn.~(\ref{criterionFour}) reproduces Eqn.~(\ref{criterionThree}) precisely. 

\bigskip

\section{More Events; Other Dimensions}

It is natural to attempt to generalize the above considerations to more than 3 events, and to spacetimes of other dimensions.  

In the case of four events in 3+1 dimensions, one can give a similar geometric criterion for when they can be placed in arbitrary time order:
Generically, four events $A,\ B,\ C,\ D$ uniquely define a three-dimensional hyperplane.  If the normal vector to this hyperplane is timelike, then a
 frame can be chosen in which the normal vector lies entirely in the time direction;  in this frame the four events are simultaneous, and small Lorentz boosts can be made to achieve any of the 24 possible time-orderings of $A,\ B,\ C,\ D$.  On the other hand, if the normal vector is spacelike then the hyperplane will contain a timelike vector and not all orderings will be realized.  To enumerate all possible cases one must also consider whether each triple of events $(A,\ B,\ C)$, $(A,\ B,\ D)$, etc., meets or fails to meet the chronology criterion of the previous section.   A more detailed analysis appears in the Appendix.   

In general, one can not order more than four events arbitrarily in 3+1 dimensions.  Indeed, with five events, the four difference vectors $s_1 = A-B,\ s_{2}=A-C$ etc. will generically span spacetime, so that their span will include timelike vectors, leading to constraints on the possible chronologies that can arise.  The problem then breaks up into a number of cases and subcases corresponding to whether each of the subsets containing 4 events or 3 events satisfies its own chronology criterion. 

This direct approach to the enumeration of all possible constraints quickly becomes unmanageable as the number of events increases.  To address that problem we can make use of a more powerful formulation due to Heiligman \cite{Heiligman} and Stanley \cite{Stanley}, based on the theory of hyperplane arrangements. 

\bigskip

\noindent{\it Hyperplane Arrangements}

\bigskip

To investigate the possible orderings of $k$ events in $(n+1)$-dimensional Minkowski space, we 
consider the set of all inertial reference frames related by a simple Lorentz boost,  parameterized by the boost velocity $v$ with respect to a fiducial reference frame.   (We ignore the effects of spatial rotations, since these do not change time orderings.)
Physical velocities are those velocities with magnitude less than the speed of light, $|v|<1$.  For the moment, we do not impose this 
restriction on the set of $v$. 

In this $n$-dimensional velocity space $\cal V$, any two events $A_i$ and $A_j$ define an $(n-1)$-dimensional hyperplane of simultaneity, which we denote $H_{ij}$, consisting of those boosts which define reference frames in which $A_i$ and $A_j$ are simultaneous.  On one side of the hyperplane, $A_i$ occurs before $A_j$;  on the other side, the ordering is reversed.  If $A_i$ and $A_j$ are spacelike separated, the hyperplane $H_{ij}$ intersects the ball of physical velocities $|v|\le 1$, which is a representation of the well-known fact that physical reference frames exist that realize either time-ordering of $A_i$ and $A_j$.   If $A_i$ and $A_j$ are timelike separated, their time-ordering can be reversed by means of a boost with $|v|>1$.   For simplicity, we ignore the case of lightlike separated $A_i$ and $A_j$.

Now consider $k$ events $A_1$, $A_{2},\cdots,$ $A_{k}$ in ${\bf R}^{1,n}$.   Each pair of events $A_{i}\ne A_{j}$ defines a hyperplane $H_{ij}$.  The $\frac{1}{2} k(k-1)$ distinct hyperplanes divide $\cal V$ up into distinct regions, corresponding to distinct time-orderings of $A_1$, $A_{2},\cdots,$ $A_{k}$.   The time-orderings realizable by physical Lorentz boosts, i.e., boosts with velocity $|v|<1$, correspond to those regions that intersect the interior of the sphere $|v|=1$.   

The intersection of two or more hyperplanes $H_{ij}$ is a hyperplane of higher codimension, along which more than two events are simultaneous.   For example, along the intersection of $H_{12}$ and $H_{34}$, $A_{1}$ and $A_{2}$ are simultaneous, and so are $A_{3}$ and $A_{4}$, but the relative ordering of the first and second pairs of events are not determined.  We denote this intersection by $H_{(12)(34)}$.    As another example, consider the intersection of $H_{12}$ and $H_{23}$, where $A_{1}$, $A_{2}$, and $A_{3}$ are simultaneous.  Hence $H_{13}$ also passes through this intersection.  We denote the intersection of these three hyperplanes by $H_{123}$.   Likewise,  the intersection of $H_{i_{1}i_{2}}\, , H_{i_{2}i_{3}}\, ,\ldots,\, H_{i_{m}i_{m+1}}$ is written as $H_{i_{1}i_{2}\ldots i_{m+1}}$.  
An obvious extension of this notation shows that higher hyperplane intersections are in correspondence with partitions of $k$ integers 1, 2, $\ldots,\ k$.  Thus, the partition $(123)(45)(6)$ corresponds to the hyperplane intersection $H_{(123)(45)(6)}$, along which $A_{1},A_{2},A_{3}$ are simultaneous and $A_{4}$ and $A_{5}$ are simultaneous. (In writing partitions, we will generally neglect blocks consisting of a single integer.)

In $n$-dimensional velocity space, any set of $m\le n$ hyperplanes $H_{ij}$ will generically have a nonempty intersection.   
Generic intersections are in one-to-one correspondence with partitions containing at least $k-n$ blocks \cite{Stanley}.
For example,  intersections of 3 hyperplanes, like $H_{123}$,  are generic in $2+1$ dimensions, since the partition $(123)(4)(5)\cdots(k)$ contains $k-2$ blocks. 
Note that it is {\it not} true that $m$ hyperplanes may intersect generically  {\it only if} $m\le n$; for example, the intersection $H_{i_{1}i_{2}\ldots i_{n+1}}$ is generic, but is actually the intersection of $\frac{1}{2} n(n+1)$ hyperplanes.

Following \cite{Stanley}, we will restrict attention to generic hyperplane arrangements, in which only generic intersections occur.  Then, close to a generic intersection $H_{(i_{1}\ldots i_{l})(j_{1}\ldots j_m)(\cdots)}$, an infinitesimal Lorentz boost may be found which places $A_{i_{1}},A_{i_{2}},\ldots,A_{i_{l}}$ in any arbitrary relative order, as well as $A_{j_{1}},$ $A_{j_{2}}, \ldots,$ $A_{j_{m}}$, and so on.   

The total number of distinct regions in velocity space has been computed usingm the combinatoric theory of hyperplane arrangements \cite{Stanley}.  The result
$$
{\cal O}(A_{1},\ldots,A_{k})= c(k,k)+c(k,k-1)+\cdots+c(k,k-n)
$$
involves the signless Stirling number of the first kind, $c(k,i)$, which counts the number of permutations of $k$ elements with $i$ distinct cycles.    $c(k,k)=1$ counts just the trivial permutation and $c(k,k-1)=\frac{1}{2} k(k-1)$ is the number of simple permutations $(ij)$. 

In general, not all of these regions will intersect the physical region $|v|<1$.   The total number of physical time-orderings is therefore bounded above by the total number of regions: 
$$
{\cal O}_{\rm phys}(A_{1},\ldots,A_{k})\le c(k,k)+c(k,k-1)+\cdots+c(k,k-n)
$$

For illustration, consider the case of three events $A_{1}$, $A_{2}$, $A_{3}$ in 2+1 dimensions.    The three lines of simultaneity $H_{12}$, $H_{13}$, $H_{23}$ intersect at a point, $H_{123}$. 
If $H_{123}$ lies inside the circle $|v|=1$ then a physical frame exists in which the three events are simultaneous.  The 3 lines divide the interior of the circle into 6 regions, corresponding to the 6 distinct time-orderings of the three events (Figure 1).   Thus all possible time-orderings are realized in this case.  
If $H_{123}$ lies outside $|v|=1$ but the interior of the circle still has nonempty intersection with the 3 lines of simultaneity, then the interior of the circle is divided into  4 regions, as shown in Figure 2.  If the circle only intersects with one line of simultaneity, then only two of the three events --  the two events associated with that line -- are spacelike separated.  In general if $H_{ij}$ lies outside of $|v|=1$, the corresponding two events are timelike separated. 

\begin{figure}[ht]
\begin{center}
\includegraphics[scale=1.2]{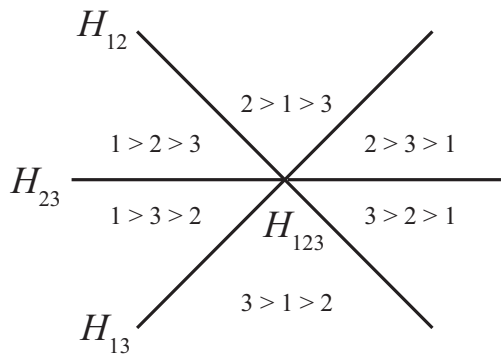}
\vspace{0.1in}
\caption{Intersection of three hyperplanes.  Along hyperplane $H_{ij}$ events $i$ and $j$ are simultaneous. Events 1, 2, and 3 are simultaneous at $H_{123}$. Time orderings of the three events in each region are indicated.}
\end{center}
\end{figure}

In $3+1$ dimensions the story for 3 events is essentially the same;  3 hyperplanes of simultaneity $H_{12}$, $H_{13}$, $H_{23}$ intersect along a line $H_{123}$. 
This case was considered in Section 1, where we found that generically all 6 orderings are realized when there is a timelike vector normal to the plane spanned by the 3 events;  otherwise at most 4 orderings could arise.  In fact the existence of such a normal vector is completely equivalent to the condition of $H_{123}$ intersecting the interior of the sphere $|v|=1$; and if $H_{123}$ lies just outside the sphere (so that $H_{12}$, $H_{13}$, $H_{23}$ still intersect the sphere and the 3 events are still spacelike separated) then generically 4 time-orderings are realized, as in Fig.~2.

\begin{figure}[ht]
\begin{center}
\vspace{0.1in}
\includegraphics[scale=1.2]{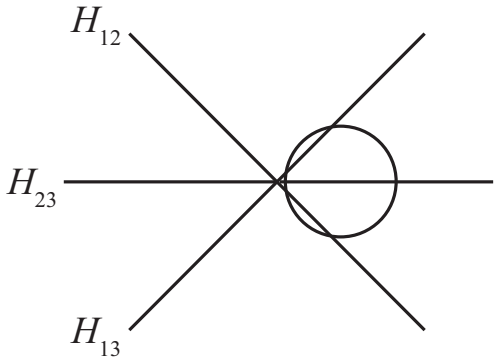}
\vspace{0.1in}
\caption{Intersection of 3 hyperplanes with the circle $|v|=1$. The physical region $|v|<1$ may intersect 1, 2, 3, 4, or 6 time-ordering regions, depending on the location of the circle. The case shown corresponds to the example of 3 events satisfying the chronology condition, as discussed in Section 1.}
\end{center}
\end{figure}

A typical configuration with four events  $A_{1}$, $A_{2}$, $A_{3}$, $A_{4}$ in $2+1$ dimensions is shown in Figure 3. $H_{12}, H_{23}, H_{34}$ bound a triangular region; the remaining 3 $H_{ij}$ bisect the 3 angles of the triangle.  Now there are a  total of six lines of simultaneity, which divide velocity space into 18 regions, each corresponding to a distinct ordering of the 4 events.  There are three double intersection points of the form $H_{(12)(34)}$, and four triple intersection points  like $H_{(123)}$.  There is an eight-parameter space of possible diagrams of the type shown in Figure 3, parameterized by the locations of the four triple intersection points in the 2D velocity space.  The eight parameters correspond to the coordinates of the four events, modulo translations and overall scalings.  As the eight moduli for the four events are varied, the locations of the triple intersection points range arbitrarily over $\cal V$.  

\begin{figure}[ht]
\begin{center}
\vspace{0.2in}
\includegraphics[scale=1.3]{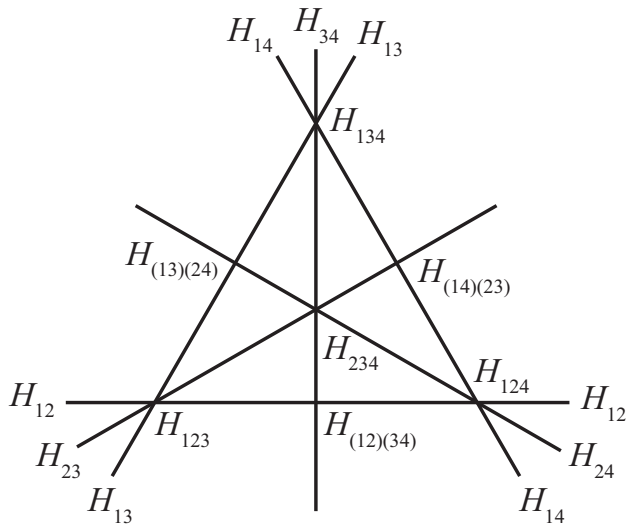}
\vspace{0.1in}
\caption{Hyperplanes of simultaneity for four events.}
\end{center}
\end{figure}

When a triple intersection point passes from the interior to the exterior of the circle, the number of distinct regions inside the circle decreases (generically) by two.  When a line or a double intersection point passes to the exterior of the circle, the number of regions intersecting the interior of the circle decreases by one.  The total number of regions inside the circle is thus equal to
$$
{\cal O}_{\rm phys}(A_{1},\ldots,A_{4})=1+N_{(ij)}+N_{(ij)(kl)}+2N_{(ijk)}
$$
where $N_{(ij)}$ denotes the number of lines of simultaneity inside the circle, $N_{(ijk)}$ is the number of triple intersection points, etc.   This formula is valid for arbitrarily many events $k$ in 2+1 dimensions. 

In $n+1$ dimensions, the above formula gets generalized to 
$$
{\cal O}_{\rm phys}(A_{1},\ldots,A_{k})=\sum_{\Pi[k]} s(H_{\Pi[k]})
$$
where the sum is over all partitions of $k$ corresponding to hyperplanes intersecting with the interior of the physical sphere $|v|=1$, and  $s(H_{\Pi})$ is a weight factor that counts the number of regions ``lost''  when such an intersection passes from the interior to the exterior of the sphere. 
The trivial partition $(1)(2)\cdots(k)$, corresponding to the whole space ${\bf R}^{1,n}$, contributes 1 to the sum. A partition with one nontrival component $(1\ldots j)(j{+}1)(j{+}2)\cdots(k)$ contributes $(j-1)!$, because when the corresponding intersection passes outside of the physical sphere, the lost regions correspond to orderings where a particular event occurs before (or after) all the others, and there are $(j-1)!$ such orderings. Note that $(j-1)!$ is equal to the number of distinct cyclic permutations associated with this partition.  It is easy to see that this fact generalizes, and  $s(H_{\Pi})$ is equal to the number of distinct cyclic permutations associated with the 
partition $\Pi[k]$. In this way, one can enumerate all the possible orderings, for any set of events in any number of dimensions, by examining all the hyperplane intersections inside of the physical sphere.   As a special case, when the physical circle is large enough that all hyperplane intersections are generic, and all of them intersect the interior of the physical sphere, the number of possible orderings is maximal and equal to
\begin{equation}\label{maxorderings}
{\cal O}(A_{1},\ldots,A_{k})= c(k,k)+c(k,k-1)+\cdots+c(k,k-n) 
\end{equation} 
recovering the result of \cite{Stanley} mentioned earlier.  The series (\ref{maxorderings}) terminates with $c(k,k-n)$ because the minimum number of blocks in a partition corresponding to a generic intersection of hyperplanes is $k-n$ \cite{Stanley}.

\section{Convexity}

In the Minkowski geometry of events, appropriate to special relativity, one studies relations among events that are invariant under Poincare transformations (translations, spatial rotations, and boosts).   An example of such an invariant relation is the time ordering among timelike and lightlike-separated events; this extends to a partial ordering on the whole of Minkowski space, with respect to which spacelike separated events are unordered \cite{KP}. This partial ordering leads to an invariant notion of ``betweenness'', according to which an event is said to occur between two timelike-separated events if it lies in the intersection of the future light-cone of the earlier event with the past light-cone of the later event. 
 
In any particular realization of Minkowski spacetime there is a natural (noninvariant) ordering of {\it all} events, given by projection onto the time coordinate.   We call betweenness relations defined with respect to such a projection chronologies.  One might have thought  that the chronologies of  spacelike separated events would be entirely frame-dependent -- as they are for two events.  But we have found that nontrivial, frame-independent chronological constraints appear even in surprisingly simple, otherwise unstructured situations. 

To place these notions in a more general context, consider any vector space $V$ with an inner product.  Then we can impose an ordering of  points in $V$, i.e. vectors, analogous to chronologies of events in Minkowski space, according to their inner products with a fixed vector.  By considering all possible such orderings, we arrive at invariant notions generalizing betweenness, analogous to our constraints on chronologies.  For up to $d+1$ generic points in $d$ dimensions there is no constraint, but for larger numbers of points constraints do arise, for reasons of convexity.  

For example, given a set of $N$ points in $V$ consider the minimal convex set containing them - their convex hull. This can be defined as the intersection of all half-spaces containing the $N$  points.  (A half-space is a set composed of all points in $V$ which are either $\ge$ or $\le$ a given point, with respect to one of the orderings defined above.) Then any point lying in the interior of the convex hull will be subject to an ordering constraint. Namely, it cannot be less than or greater than all the other points in any coordinate ordering; it must lie between events on the boundary of the hull, a sort of generalized ``betweenness"  relation.  
Further constraints are obtained by considering subsets of the original $N$ points. 

A more general condition tells us whether or not an ordering can be found for which a point $P$  is less than all of $A_1, ... , A_m$ but greater than all of $B_1, ... , B_n$.   Such an ordering exists if and only if the intersection of the convex hull of $\{ A_1, ... , A_m, P \}$ and the convex hull of $\{ B_1, ... B_n, P \}$ is simply $\{ P \}$.  By well-known hyperplane separation theorems \cite{Minkowski}, this condition is equivalent to the existence of a hyperplane containing $P$, separating the convex hull of $\{ A_1, ... , A_m \}$ from the convex hull of $\{ B_1, ... , B_m \}$.  The coordinate normal to such a hyperplane produces the desired ordering. 

Now specializing to Minkowski space, we can formulate an analogous condition for the existence of time-orderings in terms of appropriate notions of convexity.  Define the 
future (past) convex hull of a set of events to be the convex hull of the future (past) light cones of those events. The future convex hull of a set of events may also be described as the intersection of the futures of all spacelike hyperplanes whose futures contain all the events in the set.  Future and past convex hulls are invariant under changes of frame, and can be used to define invariant ordering relations.  For example, a point in the interior of the past convex hull of a set of events will not occur after all of those events in any reference frame.  

More generally, a necessary and sufficient condition that there be a chronology in which $\{ A_1, ... , A_m\}$ occur after  $\{ B_1, ... , B_n\}$, is that the intersection of the future convex hull of $\{ A_1, ... , A_m\} $ with the past convex hull of $\{ B_1, ... , B_n\} $ be empty.  This condition guarantees that a hyperplane separating the two convex sets exists.  Such a hyperplane is automatically spacelike; if it were not, then it would contain timelike rays, which would intersect any light cone in the future or past convex hulls.  The coordinate normal to such a hyperplane is thus timelike, and produces the desired time-ordering. 

The picture in velocity space is as follows.  The condition that $\{ A_1, ... , A_m\} $ occur before $\{ B_1, ... , B_n\} $ is equivalent to $mn$ conditions $A_{i}^{t}<B_{j}^{t}$, and is thus 
associated with a region in $\cal V$ bounded by $mn$ hyperplanes $H_{A_{i}B_{j}}$.  This convex region is the intersection of the half-spaces in $\cal V$ corresponding to each of the conditions  $A_{i}^{t}<B_{j}^{t}$.  In any frame corresponding to a point in this region, the conditions  $A_{i}^{t}<B_{j}^{t}$ are automatically satisfied, and there is a constant-time hypersurface separating the $A$'s from the $B$'s.   Thus, if and only if this region intersects the unit sphere, a physical frame exists in which an ordering of the desired type is achieved.

To close this section let us reconsider our original example of three spacelike separated points, $A$, $B$, and $C$, in this more general context, and show how the separation theorem of the preceding paragraph reproduces  the chronology condition (\ref{criterionOne}).  Suppose the ordering $A_t < B_t < C_t$ is impossible.  Then (after bringing $B$ to the origin by a translation), the convexity criterion states that the convex body spanned by the  past light cones of $A-B$ and the origin intersects the convex body spanned by the future light cones of $C-B$ and the origin nontrivially.  Let $v$ be an event in the intersection; we then have 
\begin{eqnarray}
v \, &=& \, \lambda_1 ( (A-B) +  l_1) + (1-\lambda_1 ) l^\prime_1 \\
\, &=& \, \lambda_2 ( (C-B) + l_2  ) + (1 -\lambda_2 ) l^\prime_2 
\end{eqnarray} 
where $l_1, l^\prime_1, l_2, l^\prime_2$ are lightlike or timelike vectors with $l_{1t}, l^\prime_{1t} <  0$, $l_{2t}, l^\prime_{2t} >0$, and $0 \leq \lambda_1, \lambda_2 \leq 1$.   It follows that 
\begin{equation}
\lambda_1 (A-B) - \lambda_2 (C-B) \, = \, \lambda_2 l_2 - \lambda_1 l_1 + (1 -\lambda_2 ) l^\prime_2  - (1-\lambda_1 ) l^\prime_1
\end{equation}
is lightlike or timelike, since all terms on the right side are future-directed lightlike or timelike vectors; thus the chronology criterion is met.   By the same reasoning, if the chronology criterion is unmet then the relevant intersections will be trivial for all orderings, and each ordering will be achieved in some frame.

\bigskip

\section{Metric from Chronology Assignments}

Thus far we have regarded the spacetime metric as given, and studied the chronologies it supports.   To round out our discussion, we will now ask a sort of converse: Given knowledge of all allowed chronologies, to what extent can we reconstruct the metric?  (First, a disclaimer: Extraction of metric data from causal data has been one of the central programs of the causal set approach to quantum gravity and its cousins \cite{Sorkin}, where in particular it has been shown that the metric can be reconstructed, up to a conformal factor, from a knowledge of the ordering relationships between timelike and lightlike separated points.  Indeed, the program of reconstructing geometry from causality dates back at least to the Robb's work of 1914 \cite{Robb}.    We are aware that in the first part of this Section we are essentially rederiving a known result from a special perspective.)

To be more precise, the mathematical form of our problem is as follows.   We want to consider spacetimes that support Lorentz symmetry.   This means that our spacetime comes equipped with a quadratic form $h_{\mu \nu}$ of signature $(1,3)$.   $h_{\mu \nu}$, regarded as a matrix, is related to the usual Minkowski metric $\eta \equiv {\rm diagonal~} (1, -1, -1, -1)$ by a similarity transformation
\begin{equation}
\eta \, = \, S^{\rm T} h S
\end{equation}
corresponding to a GL$(4,{\bf C})$ transformation
\begin{equation}
x \, \rightarrow \, Sx 
\end{equation}
on the spacetime.  Then $h$ is invariant under an SO$(1, 3)$ group of transformations of the form 
\begin{equation}
h \, = \, (SUS^{-1}) h (SUS^{-1})^{\rm T}
\end{equation}
Here $U$, obeying $\eta = U \eta U^{\rm T}$, is a Lorentz transformation in the standard form.   Our problem is to reconstruct, from the chronologies our spacetime supports, its particular $h$.   We shall show that it is always possible to do this, up to an overall constant; that is obviously the most that could be hoped for.   

Our reconstruction of the metric will employ chronology conditions to identify, by self-consistency requirements, the coordinates of events in a frame where the metric is diagonal and of the form  $$\eta = \lambda~{\rm diagonal} (1, -1, -1, -1).$$  The reconstruction proceeds by a four-step process: 
\begin{enumerate}
\item A maximal set of simultaneity is a set of events that all occur at the same time in some frame, but is not contained in any larger set of that kind.   Maximal sets of simultaneity are surfaces of constant time in some frame.  Pick such a set.   
\item The maximal sets of simultaneity that do not intersect the set chosen in Step 1 are the other surfaces of constant time in the frame chosen in Step 1.   In that frame, or in any frame that leaves the same set of maximal sets of simultaneity invariant, the chronological ordering among the surfaces is fixed by the chronology of any events within the corresponding surfaces.  So now we have a foliation by space-like hyperplanes, and an ordering of those surfaces, but not yet any metric structure.
\item Let $P$ be an event in one time hyperplane, which we give the coordinate $\tau = 0$.   The solid future light-cone of $P$ can be defined purely chronologically, as the set of events that occur after $P$ in all chronologies; similarly the solid past light-cone of $P^\prime$ is the set of events that occur before $P^\prime$ in all chronologies. Choose another time hyperplane, coming after $\tau =0$ in the chosen frame, and label it $\tau = \tau_0$.    Now consider the projections of past solid light-cones emanating from events $P^\prime$ on constant-time hyperplanes between $\tau = 0, \tau_0$, and compare them with the projection of the future solid light-cone of $P$.   There will be exactly one point $P^\prime$, and one hyperplane, for which these projections coincide.   That will occur when $P^\prime$ occurs at the same spatial coordinate as $P$ (in the preferred frame) and for the hyperplane whose time is the average $\tau_0/2$ (in the preferred frame).     We can iterate the process, finding hyperplanes with $\tau = k \tau_0/2^n$, $0 \leq k \leq 2^n$, and points on these planes that have the same spatial position.   We can also extend the construction by bringing in hyperplanes outside the interval $\tau = 0, \tau_0$ in the construction of this Step.   Thus we can construct world-lines of stationary observers in the chosen frame; and establish the appropriate measure of time-intervals on each such world-line.
\item Finally we can construct the spatial distance-function in each hyperplane, as follows.    Let $P$ and $P^\prime$ be points on the same stationary-observer world-line, as constructed in Step 3, and let the time interval between them be $\tau_1$.   If $Q$ is an event in the hyperplane of $P^\prime$, we assign spatial distance $\tau_1$ between $Q$ and $P^\prime$ if $Q$ is in the solid future light-cone of $P$, but not in the solid future light-cone of any event along the stationary-observer world-line lying in the interval between $P$ and $P^\prime$.   In this way, varying $P$ and $P^\prime$, we can determine the spatial distance between any two points.   
\end{enumerate}
Thus considerations of chronology allow us to identify a coordinate system in which the metric is diagonal and of the form $\eta = \lambda\>{\rm diagonal~} (1, -1, -1, -1)$.   The only arbitrariness arose in the choice of $\tau_0$; due to that circumstance, the metric has been determined up to an overall constant.

Since any Lorentzian spacetime can be approximated locally by \hyphenation{Min-kow-ski} Minkowski spacetime, and our arguments are essentially local, our result can be taken over to that more general context.  In each local patch, we consider time coordinates that match the local flat-space time, i.e. flows generated by approximate time-like Killing vector fields.  Then we've shown that chronology conditions determine the metric up to a conformal factor, i.e. a position-dependent scale.   

It is possible that more detailed considerations, quantifying the notion of approximate timelike Killing vector fields and paying careful attention to the (competing) effects of curvature in the small, would allow a complete reconstruction of metric from chronology, up to an overall constant, even in a curved spacetime.   Leaving that possibility to future work, in the next Section we will consider a looser extension of chronology relations to curved spacetimes, allowing for a much wider class of time variables.

\bigskip

\section{Curved Spacetimes}

So far we have considered flat spacetime and, as is natural in that context, only allowed time slicings which respect the symmetry of flat spacetime.    When we pass to the curved spacetimes of general relativity, it is natural to allow for more flexible definitions of time coordinates.   We only require that the surfaces of constant ``time'' be spacelike, i.e. that their tangent vectors are all spacelike, or equivalently that their normal vectors are timelike.   In this context the notion of convexity, which was central in flat spacetime, falls away.   Indeed even the bare possibility of introducing monotonic chronologies is not guaranteed for arbitrary manifolds of Lorentz signature; famously, one can encounter closed timelike loops.     In order to find a recognizable generalization of our flat-space result, we focus on a broad class of spacetimes that has long been considered in connection with questions involving causality, the {\it strongly causal\/} spacetimes \cite{HawkingEllis}.   For our purposes, we can take as the defining property of a strongly causal spacetime a condition that is usually derived from (and proved equivalent to) other, more abstract conditions, to wit: One can define on the spacetime a function $\tau$ whose level surfaces are spacelike.  

We define the causal future of a set $A$ to be the set $J^+(A)$ of all points that can be reached by a future-directed non-spacelike curve emanating from a point in $A$. In Minkowski space, $J^+(A)$ is the union of all future light cones, with their interiors included, emanating from $A$.  Similarly we define the causal past $J^-(B)$ to be the set of all points that can be reached by a past-directed non-spacelike curve emanating from a point in $B$.    Note that these are natural generalizations of the convex sets of Section 3 to curved spacetimes;  $J^+(A)$ may alternatively be defined to be the intersection of the futures of all spacelike hypersurfaces  whose futures contain $A$. 
We then assert: In a strongly causal spacetime, there is a time slicing in which every event in a set $A = \{ A_1 ,  ... , A_ m \}$ occurs after every event in another set $B =  \{ B_1, ... , B_n \}$ if and only if $J^+(A)$ is disjoint from $J^-(B)$. Necessity is obvious; we now indicate how one might demonstrate sufficiency.  

Our construction has three parts.   First, we define a particular spacelike surface $S$, partly adapted to the geometry of $J^+(A), J^-(B)$ and partly adapted to the foliation defined by $\tau$, which separates $J^+(A)$ from $J^-(B)$.   Second, we embed $S$ as a level surface of the flow of a quasi-time variable $\tilde \tau$.   The level surfaces of $\tilde \tau$ are spacelike, and its flow covers the entire spacetime.   Our $\tilde \tau$ is not quite a legitimate time variable, however, because an event can correspond to an interval of $\tau$, rather than a single value.   The third step, which we do not carry out, would be to perturb $\tilde \tau$ into a legitimate time variable.   While we think it is highly plausible that this can be done, we have not proved it.   

Let $\tau_2$ be the smallest value that $\tau$ assumes on $A$, $\tau_1$ the largest value that $\tau$ assumes on $B$, and $\tau_0$ the smallest value that $\tau$ assumes on $B$.   (Note that it is possible to have $\tau_2 < \tau_0$.)  Let $\tau_{\hbox{\tiny $S$}}$ be less than $\tau_0, \tau_1, \tau_2$.   Then we define $S^\prime$ to be the frontier of the union of the set of events with $\tau \leq \tau_{\hbox{\tiny $S$}}$ and $J^-(B)$; thus over parts of its range $S^\prime$ will be the $\tau = \tau_{\hbox{\tiny $S$}}$ plane, and over other parts it will coincide with $J^-(B)$.   $S^\prime$ is a non-timelike surface with $J^+(A)$ on its future side and $J^-(B)$ in its past or present.  By a small deformation we can promote it to a spacelike surface $S$ that separates $J^+(A)$ from $J^-(B)$ (and $A$ from $B$).   

We begin to define a foliation of surfaces associated with the variable $\tilde \tau$ as follows:
\begin{itemize}
\item For $\tilde \tau \leq \tau_{\hbox{\tiny $S$}}$, take the surface $\tau = \tilde \tau$.
\item For $\tau_{\hbox{\tiny $S$}} \leq \tilde \tau \leq \tau_1$, take the surface defined by those events in $S$ with $\tau \leq \tilde \tau$ and those events in the past of $S$ with $\tau = \tilde \tau$.
\end{itemize}
We can describe the resulting $\tilde\tau$ flow from the point of view of a local observer whose timelike worldline follows the gradient flow of the function $\tau$.  Suppose that the observer carries both a $\tau$ and a $\tilde \tau$ clock with him.  Prior to reaching the surface $S$, the $\tilde \tau$ and $\tau$ clocks coincide. Upon hitting $S$,
the $\tilde\tau$ clock continues running while the $\tau$ clock remains fixed, until all the worldlines in the spacetime have reached $S$.   The $\tilde \tau = \tau_1$ surface is $S$.    

If $\tau_1 < \tau_2$ we can complete the construction of $\tilde \tau$ surfaces according to the following procedure:
\begin{itemize}
\item For $\tau_1 \leq \tilde \tau \leq 2 \tau_1 - \tau_{\hbox{\tiny $S$}}$, take the hypersurface composed of those events in the future of $S$ with $\tau = \tilde \tau - \tau_1 + \tau_{\hbox{\tiny $S$}}$ and those events in $S$ with $\tau \geq \tilde \tau - \tau_1 + \tau_{\hbox{\tiny $S$}}$.
\item For $\tilde \tau \geq 2 \tau_1 - \tau_{\hbox{\tiny $S$}}$, take the surface $\tau = \tilde\tau - \tau_1 + \tau_{\hbox{\tiny $S$}}$.  
\end{itemize}
In words: As the $\tilde\tau$ clock proceeds on its worldline into the future of $S$, it resumes running at the same rate as the $\tau$ clock, with an offset for the maximum shift in $\tilde \tau$ compared to $\tau$ (the time interval $\tau_1-\tau_2$ over which the $\tilde\tau$ clock runs while $\tau$ is fixed)
introduced in the preceding steps.  The flow resumes immediately for worldlines that reached $S$ first, and is phased in as $\tilde\tau$ advances.  Finally when all worldlines have running $\tau$-clocks, we simply follow the (offset) $\tau$ flow.

\begin{figure}[ht]
\begin{center}
\includegraphics[scale=0.6]{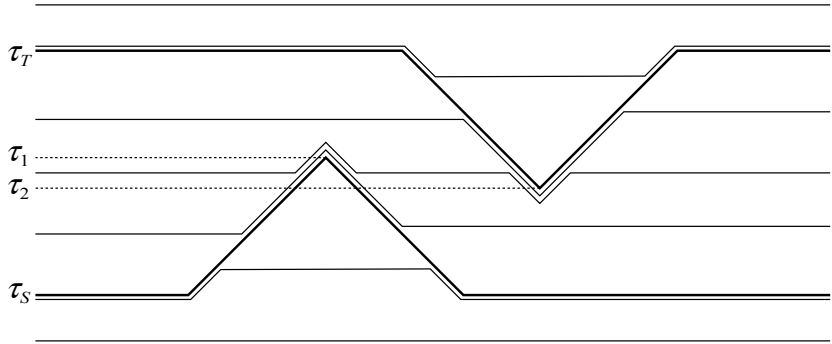}
\caption{Construction of a time foliation $\tilde\tau$ with respect to which all events in set $A$ occur after all events in set $B$.
For simplicity, we depict the case where $A$ and $B$ each consist of a single event. 
$\tau_1$ is the maximum value $\tau$ assumes over set B;  $\tau_2$ is the minimum value of $\tau$ on set A. Thick lines are surfaces $S$ and $T$, as described in text.  
Thin lines are surfaces of constant $\tilde\tau$.}
\end{center}
\end{figure}

Of course, for $\tau_1 < \tau_2$ we could have simply used the original $\tau$ surfaces; but the preceding construction is part of a more general method.  If $\tau_1 > \tau_2$ another step is required.   One introduces an additional auxiliary spacelike surface $T$, analogous to $S$, but based on a large $\tau_{\hbox{\tiny $T$}}$ and $J^+(A)$ instead of $\tau_{\hbox{\tiny $S$}}$ and $J^-(B)$.  One then constructs a $\tilde \tau$ flow that first adapts to $S$, as before, and then leaves $S$ and adapts to $T$. 
In this second, intermediate phase, a surface of constant $\tilde\tau$ consists of those events between $S$ and $T$ with $\tau = \tilde \tau - \tau_1 + \tau_{\hbox{\tiny $S$}}$, those events in $S$ with $\tau \geq \tilde \tau - \tau_1 + \tau_{\hbox{\tiny $S$}}$, and events in $T$ with $\tau \leq \tilde \tau - \tau_1 + \tau_{\hbox{\tiny $S$}}$.  When $\tilde\tau= \tau_{\hbox{\tiny $T$}}-\tau_{\hbox{\tiny $S$}}+\tau_1$, the $\tilde\tau$ surface coincides with $T$, and the  $\tilde\tau$ flow resumes until it rejoins (with a constant shift) the original $\tau$ flow.  The $\tilde\tau$ flow thus constructed is depicted in Figure 4, for the special case where the sets $A$ and $B$ each consist of one event. 

\bigskip
 
We thank Ben Braun, Jeffrey Goldstone, and Oleg Lunin for useful discussions. AS is supported by National Science Foundation Grants PHY-0555444 and PHY-0855614.  FW is supported in part by the U.S. Department of Energy under contract No. DE-FG02-05ER41360.

\section*{Appendix: Four Events, Analytically}

Consider four spacelike separated events $A$, $B$, $C$, $D$, and the vectors $v_1 = A-D$, $v_2 = B-D$, $v_3 = C-D$.  Inspired by the geometric formulation of the three-event case, we consider the normal vector $n\equiv *(v_1\wedge v_2 \wedge v_3)$ and analyze what happens when it is timelike or spacelike.  (From now on we will, for simplicity of exposition, suppose that our events are in generic positions, unless stated otherwise.)  

Note that if $n$ is timelike then all vectors in the hyperplane spanned by $v_1, v_2, v_3$ are spacelike, whereas if $n$ is spacelike then the span of $v_1, v_2, v_3$ includes timelike vectors.  Indeed, if $n$ is timelike, all vectors orthogonal to $n$ are spacelike, and these are precisely the vectors in the span of  $v_1, v_2, v_3$.  Conversely, if $n$ is spacelike, its orthogonal complement includes timelike vectors.
The claim being covariant, we can check it in any convenient frame.    If $n^\mu \equiv \epsilon^\mu_{\ \nu\rho\lambda}v_1^\nu v_2^\rho v_3^\lambda$ is timelike we can choose a frame wherein $n \propto (1, 0, 0, 0)$; in this frame it is obvious that vectors orthogonal to $n$ are spacelike, and since all the vectors spanned by $v_1, v_2, v_3$ are orthogonal to $n$, the desired conclusion follows.   If $n$ is spacelike we can choose a frame with $n \propto (0, 1, 0, 0)$; then it is clear that the hyperplane orthogonal to $n$, which is the span of $v_1, v_2, v_3$,  contains timelike vectors, {\it e.g.} $(1,0, 0 ,0)$.

We now claim that if $n$ is timelike, then all possible chronologies among $A, B, C, D$ are possible (similarly to our earlier Case Two).    To see this, begin again with the frame in which only the time component of $n$ is nonzero.  We argue that suitable small perturbations from this frame can realize any desired chronology.  For definiteness, consider the chronology $A^t <  B^t < C^t < D^t$.   We can choose axes so that  $v_1=(0,v_1^x,0,0)$, $v_2=(0,v_2^x,v_2^y,0)$, and $v_3=(0,v_3^x,v_3^y,v_3^z)$ .   An infinitesimal boost in the $x$ direction will serve to make the time component of $v_1$ negative; following this, a (rather less) infinitesimal boost in the $y$ direction will make the time component of $v_2$ more negative, without changing the sign of $v_1$; a final boost in the $z$ direction will make  $v_3$ most negative, without changing the signs or relative magnitudes of $v_1$ and $v_2$.   Thus the desired chronology is achieved.

Next consider the case where $n$ is spacelike. 
We have shown above that in this case there must be a linear combination of $v_1,v_2,v_3$ that is timelike.   We have then a relation of the form
\begin{eqnarray}\label{4EventsCondition}
aA + bB + cC + dD \, &=& \, {\rm timelike} \nonumber \\
a + b + c + d \, &=& \, 0 \, .
\end{eqnarray}
Without significant loss of generality we may suppose that the time component of $aA + bB + cC + dD$ is positive and that either $a, b, c > 0$ and $d <0$, or $a,b > 0$ and $c,d < 0$.   In the first case, it is impossible for $D$ to come after all of $A, B, C$.   In the second case, it impossible for both of $C, D$ to come after both $A, B$, i.e., the chronologies $A^t < B^t < C^t < D^t$, $A^t < B^t < D^t < C^t$, $B^t < A^t < C^t < D^t$, and $B^t < A^t < D^t < C^t$ are forbidden.   

An analogue of the chronology criterion (\ref{criterionOne}) may be obtained by 
squaring $n^\mu \equiv \epsilon^\mu_{\ \nu\rho\lambda}v_1^\nu v_2^\rho v_3^\lambda$:
\begin{equation}\label{4EventsCriterion}
n^2 = \epsilon_{\mu\alpha\beta\gamma} \epsilon^{\mu}_{\ \nu\rho\lambda}\, v_1^\alpha v_2^\beta v_3^\gamma v_1^\nu v_2^\rho v_3^\lambda \le 0 \, .
\end{equation}
Expanding the epsilons out in terms of inner products would give the 4-event generalization of Eqn.~(\ref{criterionOne}). 
We may also write Eqn.~(\ref{4EventsCriterion}) in terms of $A$, $B$, $C$, $D$ as 
\begin{equation}
(A\wedge B \wedge C - A \wedge B \wedge D + A \wedge C \wedge D - B \wedge C \wedge D)^2 \le 0
\end{equation}
analogous to Eqn.~(\ref{criterionFour}). 

\bigskip
The preceding analysis is certainly not exhaustive.   Specifically, it can occur that just three events already support a relationship of the kind (\ref{interpretationTwo}), in which case the analysis of Section 1 applies, and the forbidden chronologies among three events remain forbidden, regardless of when the fourth occurs.   In this case, one has in effect Eqn.~(\ref{4EventsCondition}) with $d=0$.   Note that the left-hand side will remain timelike with $d$ infinitesimally positive or negative.    More generally, one can have Eqn.~(\ref{4EventsCondition}) holding true for a range of parameters, including differently signed combinations of $a$, $b$, $c$, $d$, resulting in additional constraints.   A complete classification appears complicated, and we have not seriously attempted it. 

\bigskip
Generically, one can not order five or more events arbitrarily.  Indeed, the four difference vectors $v_1 = A-B$ etc. will span spacetime, so that a solution to 
\begin{eqnarray}
aA + bB + cC + dD + eE \, &=& \, (1, 0, 0, 0)  \nonumber \\
a + b + c + d + e \, &=& \, 0
\end{eqnarray}
will always exist.  Constraints arise from this relation.  For example, if (say) $a, c, e > 0$ and  $b, d <0$, then there are no chronologies in which both $B$ and $D$ follow every one of $A, C, E$.

\end{document}